\documentstyle[12pt]{article}
\begin{document}
\baselineskip=7mm
\def\ps{p\kern-0.5em/\kern0.0em}

\noindent
January 1997 (revised April 1997), SAGA-HE-112-97

\bigskip

\centerline{\bf Meson-nucleon vertex corrections to the  vacuum polarization}
\centerline{\bf in the cutoff field theory and tensor coupling}

\bigskip

\centerline{By}

\noindent{\bf Hiroaki Kouno$^*$, 
Kazuharu Koide, Nobuo Noda, Katsuaki Sakamoto, 
}

\noindent{\bf
Yoshitaka Iwasaki,
Tomohiro Mitsumori, 
 Akira Hasegawa
}

\centerline{Department of Physics, Saga University, Saga 840, Japan}

\centerline{and}

\centerline{\bf Masahiro Nakano}

\centerline{University of Occupational and Environmental Health, Kitakyushu 807, Japan}

~

\noindent
*e-mail address: kounoh@cc.saga-u.ac.jp

\bigskip

\centerline{\bf PACS numbers: 21.65.+f}

\bigskip

\bigskip

\centerline{\bf Abstract}

\bigskip

\noindent
At zero-density, meson-nucleon vertex corrections to vacuum polarization are studied in the $\sigma$-$\omega$ model with the cutoff. 
It is shown that the properties of the vertex corrections to vacuum polarization are somewhat different from those described by the ordinary renormalization procedures when the cutoff is small ( $< 5$GeV). 
The low-energy effective Lagrangian is constructed in the framework of the renormalization group method. 
The weak tensor and derivative couplings of meson-nucleon interactions may be needed in the low-energy effective theory of mesons and nucleons. 

\vfill\eject



\centerline{\bf 1. Introduction}

\bigskip

Nuclear matter has been studied in the framework of quantum hadrodynamics (QHD) in the recent two decades. 
The meson mean-field theory for nuclear matter [1] has made successful results to account for the saturation properties at the normal nuclear density. 
Following to those successes, many studies and modifications are done in the relativistic nuclear models. 
One of those modifications is inclusion of vacuum fluctuation effects, which cause divergences of physical quantities as they are naively calculated. 
Chin [2][3] estimated the vacuum fluctuation effects in the Hartree approximation removing the divergences by the renormalization procedures, and found that the vacuum fluctuation effects make the incompressibility of nuclear matter smaller and closer to the empirical value than in the original Walecka model. 
However, it becomes more difficult to do the renormalization as the model becomes more complicated, since the renormalization procedures need analytical studies to some extent. 
Numerical studies based on the cutoff field theory may be useful. 

Furthermore, the relation between QHD and the underlying fundamental theory, i.e., QCD, is the open question. 
One may wonder whether QHD is valid in very high-energy scale or not. 
If QHD is valid only under some energy scale, it is natural to introduce the cutoff or the form factor into the theory. 
One may introduce the cutoff [4] or the form factor [5] to avoid the instability of the meson propagators  in the random phase approximation (RPA) [6]. 
Cohen [7] introduced the four dimensional cutoff into the relativistic Hartree calculation and found that the vacuum energy contribution is somewhat different from the one in the renormalization procedures, if the cutoff is not so large. 

In the recent paper [8], we have studied the nuclear matter properties in details using the cutoff field theory of Cohen and show that the properties of nuclear matter may be somewhat different from those described in the ordinary renormalization procedures. 
In the same paper, the cutoff field theory of the nuclear matter is reformulated in the framework of the renormalization group methods [9][10][11] and the fifth and sixth order terms of $\sigma$-meson self-interactions in the effective Lagrangian are shown to be important in the cutoff field theory. 
Using the effective Lagrangian, the compressional properties of the nuclear matter are studied in detail [12]. 
We have also studied the meson masses at finite baryon density in the framework of the cutoff field theory and renormalization group methods [13]. 

In this paper, we studies the meson-nucleon vertex corrections at zero-density in the framework of the cutoff field theory. 
The results are reformulated in the framework of the renormalization group methods, and the roles of the tensor coupling and the derivative couplings in the effective Lagrangian are studied. 
This paper is organized as follows. 
In section 2, the vertex corrections in the cutoff field theory are formulated and the numerical results are shown comparing them with the results which are obtained by the ordinary renormalization procedures. 
In section 3, the vertex corrections in the cutoff field theory are reformulated in the framework of the renormalization group methods. The tensor coupling and the derivative couplings are derived in the low-energy effective Lagrangian. 
The role of the new interactions and the convergence of the expansion of the effective Lagrangian are discussed there. 
The section 4 is devoted to the summary and discussions.


\bigskip

\bigskip

\centerline{\bf 2. Vertex Corrections in the Cut-off Field Theory}

\bigskip

We start with the following Lagrangian of $\sigma$-$\omega$ model together with a regulator that truncates the theory's state space at some large $\Lambda$. 
$$
L={\bar{\psi}}(i\gamma_\mu\partial^\mu-M+g_s\phi-g_v\gamma_\mu V^\mu )\psi
$$
$$
+{{1}\over{2}}\partial_\mu\phi\partial^{\mu}\phi-{{1}\over{2}}m_s\phi^2
-{{1}\over{4}}F_{\mu\nu}F^{\mu\nu}+{{1}\over{2}}m_v^2V_\mu V^\mu
+\delta L, \eqno{(1)} $$
where $\psi$, $\phi$, $V_\mu$, $M$, $m_s$, $m_v$, $g_s$, and $g_v$ are nucleon field, $\sigma$-meson field, $\omega$-meson field, nucleon mass, $\sigma$-meson mass, $\omega$-meson mass, $\sigma$-nucleon coupling, and $\omega$-nucleon coupling, respectively. 
The term $\delta L$ is needed to make $M$, $m_s$, $m_v$, $g_s$ and $g_v$ be the physical values of the masses and the couplings. 
As far as the vertex corrections are concerned, it is given by 
$$ \delta L_{vertex} =(g_{s0}-g_s){\bar{\psi}}\phi\psi-(g_{v0}-g_v){\bar{\psi}}\gamma_\mu V^\mu\psi $$
$$=-\zeta_sg_s{\bar{\psi}}\phi\psi+\zeta_v g_v{\bar{\psi}}\gamma_\mu V^\mu\psi, \eqno{(2)} 
$$
where $\zeta_s$ and $\zeta_v$ are constant parameters which are determined phenomenologically. 
The constants $g_{s0}$ and $g_{v0}$ are called "bare" $\sigma$-nucleon and $\omega$-nucleon coupling, respectively. 
The terms such as $\delta L$ are usually called "counter terms" in the ordinary renormalization procedures. 

In fig. 1, some low order contributions of the meson-nucleon vertex correction denoted by $V(p_a,p_b)$ are shown graphically. 
The correction $\Gamma (p_a,p_b)$ to the bare meson-nucleon vertex  
is given by
$$     V_s(p_a,p_b) =(1-\zeta_s)I+\Gamma_s(p_a,p_b)    \eqno{(3)}  $$
or 
$$     V_v^\mu (p_a,p_b) =(1-\zeta_v)\gamma^\mu+\Gamma_v^\mu(p_a,p_b),    \eqno{(4)}  $$
respectively, where $I$ is a $4\times 4$ unit matrix. 


\centerline{$\underline{~~~~~~~}$}

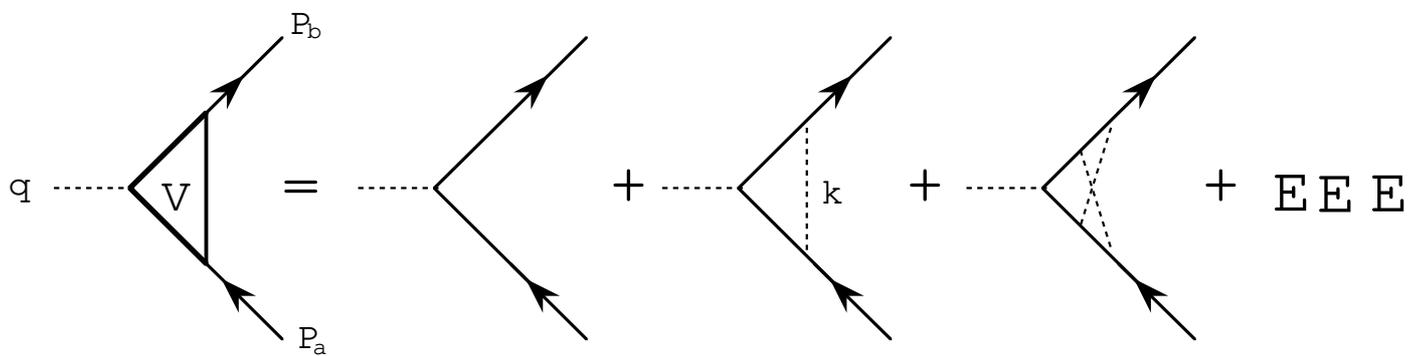
\centerline{Fig. 1}

\centerline{$\underline{~~~~~~~}$}


According to the Feynman rule, the lowest order contributions to $\Gamma_s(p_a,p_b)$ and $\Gamma_v^\mu (p_a,p_b)$ are given by [14][15]
$$ \Gamma_s(p_a,p_b)=ig_s^2\int {{d^4k}\over{(2\pi )^4}}G^0(p_b-k)G^0(p_a-k)\Delta^0 (k)  $$
$$ +ig_v^2\int {{d^4k}\over{(2\pi )^4}}\gamma^\nu G^0(p_b-k)G^0(p_a-k)\gamma^\lambda D^0_{\nu\lambda}(k)  \eqno{(5)} $$
and 
$$ \Gamma^\mu_v(p_a,p_b)=ig_s^2\int {{d^4k}\over{(2\pi )^4}} G^0(p_b-k)\gamma^\mu
 G^0(p_a-k)\Delta^0 (k)  $$
$$ +ig_v^2\int {{d^4k}\over{(2\pi )^4}}\gamma^\nu G^0(p_b-k)\gamma^\mu
 G^0(p_a-k)\gamma^\lambda D^0_{\nu\lambda}(k), \eqno{(6)}  $$
where $G^0(p)$, $\Delta^0 (p)$ and $D^0_{\mu\nu}(p)$ are the free nucleon, the free $\sigma$-meson and the free $\omega$-meson propagators, respectively. 
At zero-density, these propagators are given by 
$$    G^0(p)={{\ps+M}\over{p^2-M^2+i\epsilon}}, \eqno{(7)} $$
$$    \Delta^0 (p)={{1}\over{p^2-m_s^2+i\epsilon}} \eqno{(8)} $$
and
$$    D^0_{\mu\nu}(p)=\big(-g_{\mu\nu}+{{p_\mu p_\nu}\over{m_v^2}}\big){{1}\over{p^2-m_v^2+i\epsilon}}.  \eqno{(9)} $$

The contributions of eqs. (5) and (6) should dominate the vertex structure at large distance, since the virtual intermediate state is the one with the lowest mass [14]. 
In our pictures, the Lagrangian (1) is valid only at the momentum region which is smaller than the cutoff $\Lambda$. 
So our main interest is restricted on the small momentum region. 
Therefore, we use the eqs. (5) and (6) as the first approximations for the vertex corrections. 

The integrals in eqs. (5) and (6) diverge as they are naively calculated. 
To make the integrals finite, we introduce the Pauli-Villars regulators [16] as follows [17]. 
$$ \Gamma_s(p_a,p_b)=ig_s^2\int {{d^4k}\over{(2\pi )^4}}G^0(p_b-k)G^0(p_a-k)
\{ {{1}\over{k^2-m_s^2}}-{{1}\over{k^2-\Lambda^2}}\}
$$
$$ +ig_v^2\int {{d^4k}\over{(2\pi )^4}}\gamma^\nu G^0(p_b-k)G^0(p_a-k)\gamma^\lambda \{-g_{\nu\lambda}+{{k_\nu k_\lambda}\over{m_v^2}}\}
\{ {{1}\over{k^2-m_v^2}}-{{1}\over{k^2-\Lambda^2}}\}. \eqno{(10)} $$
$$ \Gamma^\mu_v (p_a,p_b)=ig_s^2\int {{d^4k}\over{(2\pi )^4}} G^0(p_b-k)\gamma^\mu
 G^0(p_a-k)
\{ {{1}\over{k^2-m_s^2}}-{{1}\over{k^2-\Lambda^2}}\}. $$
$$ +ig_v^2\int {{d^4k}\over{(2\pi )^4}}\gamma^\nu G^0(p_b-k)\gamma^\mu
 G^0(p_a-k)\gamma^\lambda
\{-g_{\nu\lambda}+{{k_\nu k_\lambda}\over{m_v^2}}\}
\{ {{1}\over{k^2-m_v^2}}-{{1}\over{k^2-\Lambda^2}}\}. \eqno{(11)} $$
In eqs. (10) and (11), for simplicity, we have omitted the term $i\epsilon$ in the denominators of the propagators. 
By these regulators, the integrands in eqs. (10) and (11) are suppressed in the region $\vert k^2\vert > \Lambda^2$ and the high momentum part of the integrands hardly contributes to the integrals. 
In the ordinary renormalization procedures, $\Lambda$ is taken to be infinity after the renormalization is performed. 
However, here we keep $\Lambda$ finite, since we regard $\Lambda$ as the cutoff which is the limiting energy scale of the hadronic theory. 

In general, the functions $\Gamma_s(p_a,p_b)$ and $\Gamma_v^\mu (p_a,p_b)$ have very complicated structures. 
Therefore, below we assume that the nucleons of the external lines in fig. 1 are on-shell, i.e., $p_a^2=p_b^2=M^2$. 
Under this approximation, to calculate $\Gamma_s(p_a,p_b)$ and $\Gamma_v^\mu (p_a,p_b)$ is equivalent to calculate the one-loop corrections to the amplitude for an real nucleon to scatter off an external meson field. 
It should be remarked that the term which is proportional to ${{k_\mu k_\nu}\over{m_v^2}}$ in the $\omega$-meson propagator is dropped by the baryon number conservation [3][14]. 

After some algebra and transforming the variables [17][18][14], we get 
$$ \Gamma_s(Q,\mu_L)=f_s(Q^2,\mu_L^2)$$
$$ =f_1(Q^2,\mu_s^2)-f_1(Q^2,\mu_L^2)+f_{D1}(\mu_s^2,\mu_L^2) $$
$$ +f_2(Q^2,\mu_v^2)-f_2(Q^2,\mu_L^2)+f_{D2}(\mu_v^2,\mu_L^2), \eqno{(12)} $$
and 
$$ \Gamma_v^\mu (Q^2,\mu_L^2)=f_v(Q^2,\mu_L^2)\gamma^\mu
+{{i}\over{M}}f_t(Q^2,\mu_L^2)\sigma^{\mu\nu}q_\nu $$
$$ =\{ f_3(Q^2,\mu_s^2)-f_3(Q^2,\mu_L^2)+f_{D3}(\mu_s^2,\mu_L^2) $$
$$ +f_4(Q^2,\mu_v^2)-f_4(Q^2,\mu_L^2)+f_{D4}(\mu_v^2,\mu_L^2)\}\gamma^\mu$$
$$ +{{i}\over{M}}\{ f_5(Q^2,\mu_s^2)-f_5(Q^2,\mu_L^2)
+f_6(Q^2,\mu_v^2)-f_6(Q^2,\mu_L^2)\} \sigma^{\mu\nu}q_\nu, 
\eqno{(13)} $$
where $q=p_b-p_a$, $Q^2=-q^2/M^2$, $\mu_s=m_s/M$, $\mu_v=m_v/M$, $\mu_L=\Lambda /M$ and $\sigma^{\mu\nu}={{i}\over{2}}[\gamma^\mu ,\gamma^\nu]$, respectively. 
If $Q^2\neq 0$, the functions $f_i(Q^2,\mu_s^2)~(i=1,3,5)$ and $f_i(Q^2,\mu_v^2)~(i=2,4,6)$ are given by
$$ f_1
(Q^2,\mu_s^2)
={{g_s^2}\over{16\pi^2}}\int_0^1du\big[ 
{{4(1-u)+u^2+{{Q^2u^2}\over{4}}}\over{S_s(u)Q}}
\log{ \{ {{S_s(u)+{{Qu}\over{2}}}\over{S_s(u)-{{Qu}\over{2}}}} \} }
$$
$$
-3u
+3{{S_s(u)}\over{Q}}
\log{ \big( {{ S_s(u)+{{Qu}\over{2}} }\over{ S_s(u)-{{Qu}\over{2}} }} \big) } 
\big], \eqno{(14)} $$
$$ f_2
(Q^2,\mu_v^2)
=-{{g_v^2}\over{16\pi^2}}\int_0^1du\big[ 
{{2 \{ 2(1-u+u^2)+Q^2(1-u+{u^2\over{2}})\}}\over{S_v(u)Q}}
\log{\big({{S_v(u)+{{Qu}\over{2}}}\over{S_v(u)-{{Qu}\over{2}}}}\big)}
$$
$$
-12u
+12{{S_v(u)}\over{Q}}
\log{ \{ {{ S_v(u)+{{Qu}\over{2}} }\over{ S_v(u)-{{Qu}\over{2}} }} \} } 
\big], \eqno{(15)} $$
$$ f_3
(Q^2,\mu_s^2)
=-{{g_s^2}\over{16\pi^2}}\int_0^1du\big[ 
{{-(4-u^2-{{Q^2u^2}\over{4}})+2u(2-u)}\over{S_s(u)Q}}
\log{\big({{S_s(u)+{{Qu}\over{2}}}\over{S_s(u)-{{Qu}\over{2}}}}\big)}
$$
$$
-u
+{{S_s(u)}\over{Q}}
\log{ \{ {{ S_s(u)+{{Qu}\over{2}} }\over{ S_s(u)-{{Qu}\over{2}} }} \} }
\big], \eqno{(16)} $$
$$ f_4
(Q^2,\mu_v^2)
=-{{g_v^2}\over{16\pi^2}}\int_0^1du\big[ 
{{2 \{ 2(1-u)-u^2+Q^2(1-{u\over{2}})^2\}}\over{S_v(u)Q}}
\log{ \{ {{ S_v(u)+{{Qu}\over{2}} }\over{ S_v(u)-{{Qu}\over{2}} }} \} }
$$
$$
-2u
+2{{S_v(u)}\over{Q}}
\log{ \{ {{ S_v(u)+{{Qu}\over{2}} }\over{ S_v(u)-{{Qu}\over{2}} }} \} } \big], 
\eqno{(17)} $$
$$ f_5
(Q^2,\mu_s^2)
={{g_s^2}\over{16\pi^2}}\int_0^1du
{{u(2-u)}\over{S_s(u)Q}}
\log{ \{ {{ S_s(u)+{{Qu}\over{2}} }\over{ S_s(u)-{{Qu}\over{2}} }} \} }
 \eqno{(18)} $$
and
$$ f_6
(Q^2,\mu_v^2)
={{g_v^2}\over{16\pi^2}}\int_0^1du
{{2u(1-u)}\over{S_v(u)Q}}
\log{ \{ {{ S_v(u)+{{Qu}\over{2}} }\over{ S_v(u)-{{Qu}\over{2}} }} \} }
, \eqno{(19)} $$
where $S_s(u)=\sqrt{ u^2+{{Q^2u^2}\over{4}}+\mu_s^2(1-u)}$, $S_v(u)=\sqrt{ u^2+{{Q^2u^2}\over{4}}+\mu_v^2(1-u) }$. 
If $Q^2=0$, they are given by
$$ f_1
(0,\mu_s^2)
={{g_s^2}\over{16\pi^2}}\int_0^1du
{{ u \{ 4(1-u)+u^2 \} }\over{ u^2+\mu_s^2(1-u) }}, 
\eqno{(20)} $$
$$ f_2
(0,\mu_v^2)
=-{{g_v^2}\over{16\pi^2}}\int_0^1du
{{4u(1-u+u^2)}\over{u^2+\mu_v^2(1-u)}}, 
\eqno{(21)} $$
$$ f_3
(0,\mu_s^2)
={{g_s^2}\over{16\pi^2}}\int_0^1du
{{u\{(4-u^2)-2u(2-u)\}}\over{u^2+\mu_s^2(1-u)}}, 
\eqno{(22)} $$
$$ f_4
(0,\mu_v^2)
=-{{g_v^2}\over{16\pi^2}}\int_0^1du
{{2u\{ 2(1-u)-u^2\}}\over{u^2+\mu_v^2(1-u)}}, 
\eqno{(23)} $$
$$ f_5
(0,\mu_s^2)
={{g_s^2}\over{16\pi^2}}\int_0^1du
{{u^2(2-u)}\over{u^2+\mu_s^2(1-u)}}
 \eqno{(24)} $$
and
$$ f_6
(0,\mu_v^2)
={{g_v^2}\over{16\pi^2}}\int_0^1du
{{2u^2(1-u)}\over{u^2+\mu_v^2(1-u)}}
. \eqno{(25)} $$
The functions $f_{Di}(\mu_s^2,\mu_L^2)~(i=1,3)$ and $f_{Di}(\mu_v^2,\mu_L^2)~(i=2,4)$ are given by
$$ f_{Di}
(\mu_s^2,\mu_L^2)
={{b_ig_s^2}\over{16\pi^2}}\int_0^1du
\log{ \{ {{u^2+\mu_L^2(1-u)}\over{u^2+\mu_s^2(1-u)}} \} u }
~~~~~(i=1,3), \eqno{(26)} $$
and
$$ f_{Di}
(\mu_v^2,\mu_L^2)
={{b_ig_v^2}\over{16\pi^2}}\int_0^1du
\log{ \{ {{u^2+\mu_L^2(1-u)}\over{u^2+\mu_v^2(1-u)}} \} u }
~~~~~(i=2,4), \eqno{(27)} $$
where $b_1=-2$, $b_2=8$, $b_3=1$ and $b_4=2$. 
The functions $f_i(Q^2,\mu_L^2)(i=1\sim 6)$ are gotten if $\mu_s$ or $\mu_v$ in (14)$\sim$(25) are replaced by $\mu_L$. 
It should be remarked that $Q^2=-q^2/M^2\geq 0$ when $p_a$ and $p_b$ are on-shell. 
We also remark that the functions $f_i$ are the even functions of $Q$. 
The functions (14)$\sim$(19) depend on the momenta $p_a$ and $p_b$ only through $Q^2$. 

The constant $\zeta_s$ and $\zeta_v$ are determined by the following physical conditions 
$$     V_s(p_a,p_b) =(1-\zeta_s)I+\Gamma_s(p_a,p_b)=I~~~~~({\rm at}~q=0,~p_a^2=p_b^2=M^2) \eqno{(28)}  $$
and 
$$     V_v^\mu (p_a,p_b)=(1-\zeta_v)\gamma^\mu+\Gamma_v^\mu(p_a,p_b)=\gamma^\mu~~~~~({\rm at}~q=0,~p_a^2=p_b^2=M^2). \eqno{(29)}  $$
Since the tensor part in eq. (13) disappears at $q=0$, 
the conditions above are rewritten in the following forms
$$     \zeta_s=f_s(0,\mu_L^2)~~~~~{\rm and}~~~~~\zeta_v=f_v(0,\mu_L^2). \eqno{(30)} $$
Therefore, we get
$$     V_s(p_a,p_b) =F_s(Q^2,\mu_L^2)I~~~~~({\rm at}~p_a^2=p_b^2=M^2) \eqno{(31)}  $$
and 
$$     V_v^\mu (p_a,p_b)=F_v(Q^2,\mu_L^2)\gamma^\mu +{{i}\over{2M}}F_t(Q^2,\mu_L^2)\sigma^{\mu\nu}q_\nu~~~~~({\rm at}~p_a^2=p_b^2=M^2), \eqno{(32)}  $$
where 
$$ F_s(Q^2,\mu_L^2)=1+f_s(Q^2,\mu_L^2)-f_s(0,\mu_L^2),    \eqno{(33)} $$
$$ F_v(Q^2,\mu_L^2)=1+f_v(Q^2,\mu_L^2)-f_v(0,\mu_L^2)   \eqno{(34)} $$
and 
$$ F_t(Q^2,\mu_L^2)=2f_t(Q^2,\mu_L^2),   \eqno{(35)} $$
respectively. 
It should be remarked that $f_{Di}$ in $f_s$ and $f_v$ are canceled by the subtractions at $q=0$ in eqs. (33) and (34) since the functions $f_{Di}$ do not have $q$ dependences. 
Furthermore, $f_i(Q^2,\mu_L^2)\rightarrow 0$ if the limit $\Lambda \rightarrow \infty$ is taken as in the ordinary renormalization procedures. 
Therefore, we get 
$$  F_s(Q^2,\infty )=f_1(Q^2,\mu_s^2)+f_2(Q^2,\mu_v^2)-f_1(0,\mu_s^2)-f_2(0,\mu_v^2),    \eqno{(36)} $$
$$  F_v(Q^2,\infty )=f_3(Q^2,\mu_s^2)+f_4(Q^2,\mu_v^2)-f_3(0,\mu_s^2)-f_4(0,\mu_v^2),    \eqno{(37)} $$
and 
$$  F_t(Q^2,\infty )=2f_5(Q^2,\mu_s^2)+2f_6(Q^2,\mu_v^2)    \eqno{(38)} $$
in the limit $\Lambda \rightarrow \infty$. 
Eqs. (36)$\sim$(38) are consistent with the results which are gotten by using the dimensional regularization [14][15]. 

Before showing the numerical results, we comment on the parameters which we use. 
We put $M=$939MeV, $m_s=$550MeV and $m_v=$783MeV. 
We use $g_s$ and $g_v$ which reproduce the binding energy $a_1$=15.75MeV at the normal density $\rho_0=$0.15fm$^{-3}$ in the relativistic Hartree calculation of $\sigma$-$\omega$ model (i.e., $g_s=8.81907$ and $g_v=10.0998$). 
However, the qualitative features of our results do not change if we use the other parameter sets. 

In fig. 2, the $Q$-dependence of $F_s$, $F_v$ and $F_t$ are shown. 
In these calculations, we put $\Lambda =1$ or 2GeV as examples. 
The result in the ordinary renormalization procedures (i.e., in the case of $\Lambda = \infty $) is also shown. 
In all cases, $F_s$ and $F_v$ decrease as $Q$ increases. 
The decreases of them are large when $\Lambda$ is large. 
$F_t$ also slightly decreases as $Q$ increases. 
However, the $Q$-dependence of $F_t$ is very weak. 
On the other hand, the value of $F_t$ depends strongly on $\Lambda$. 
It seems that $F_t$ becomes large when $\Lambda$ becomes large. 


\centerline{$\underline{~~~~~~~}$}

\centerline{Figs. 2 (a), (b), (c)}

\centerline{$\underline{~~~~~~~}$}


To see the $\Lambda$-dependence of the results in detail, 
in fig. 3 (dotted lines), $F_s$, $F_v$ and $F_t$ at $QM=0.5$GeV are shown as the functions of $\Lambda$. 
For comparison, the value of each function in the limit $\Lambda\rightarrow \infty $ is also denoted by the solid line which is not the function of $\Lambda$. 
$F_s$ and $F_v$ become smaller and get closer to the limiting value at $\Lambda =\infty$ as $\Lambda$ becomes larger. 
On the other hand, $F_t$ becomes larger and gets closer to the limiting value at $\Lambda =\infty$ as $\Lambda$ becomes larger. 
In all cases, the values of the functions at $\Lambda=$20GeV are almost the same as the ones in the limit $\Lambda\rightarrow \infty$. 
From these figures, we see that the vertex corrections are somewhat different for small $\Lambda (<$5GeV) from the ones for $\Lambda =\infty$. 


\centerline{$\underline{~~~~~~~}$}

\centerline{Figs. 3 (a), (b), (c)}

\centerline{$\underline{~~~~~~~}$}



\bigskip

\bigskip

\centerline{\bf 3. Low-energy Effective Lagrangian}
\centerline{\bf and the Tensor Coupling}

In this section, we reformulate the cutoff field theory, which was described in section 2, in the framework of the renormalization group method 
[9][10][11]. 
In the renormalization group method, it is required that the physical quantities should not depend on a cutoff which is conveniently introduced in the calculations. 
It should be remarked that, in this context, the "cutoff" has no physical meaning such as the limiting energy scale of the theory. 
On the other hand, it is natural to consider that there is a finite upper limit to the energy scale in QHD. 
Below, we call such a cutoff the true cutoff. 
Lepage [11] proposed to construct the low energy effective Lagrangian if the cutoff is not much larger than the energy scale in which we are interested. 

Below we construct the low energy effective Lagrangian of $\sigma$-$\omega$ model according to Lepage's pictures. 
Since we do not know the exact value of the true cutoff $\Lambda$ which is the limiting energy scale to the theory, the cutoff $\Lambda^\prime$ which we introduce into the theory is not the same as the true cutoff $\Lambda$. 
Suppose that $\Lambda^\prime$ is smaller than $\Lambda$. 
In this case, the vacuum polarization effects are underestimated. 
The contributions in the region $\Lambda^2>\vert k^2\vert >\Lambda^{\prime 2}$ of the momentum integrations are needlessly discarded. 

We estimate the contributions which are discarded by using $\Lambda^\prime$ instead of $\Lambda$. 
In the cases of the one-loop corrections to the amplitude for an real nucleon to scatter off an external meson field, i.e., in the cases of vertex corrections described in section 2, they are given by

$$  \Delta f_s(\Lambda^2>\vert k^2\vert >\Lambda^{\prime 2})
    =f_s (Q^2,\mu_L^2 )-f_s(Q^2,\mu_L^{\prime 2} )
$$
$$
    =\{ f_{D1}(\mu_s^2,\mu_L^2)-f_1 (Q^2,\mu_L^2)+f_{D2}(\mu_v^2,\mu_L^2)-f_2(Q^2,\mu_L^2) \} 
$$
$$
    -\{ f_{D1}(\mu_s^2,\mu_L^{\prime^2} )-f_1 (Q^2,\mu_L^{\prime 2} )+f_{D2}(\mu_v^2,\mu_L^{\prime 2})-f_2(Q^2,\mu_L^{\prime 2} ) \} 
, \eqno{(39)} 
$$
$$  \Delta f_v(\Lambda^2>\vert k^2\vert >\Lambda^{\prime 2})
    =f_v (Q^2,\mu_L^2 )-f_v(Q^2,\mu_L^{\prime 2} )
$$
$$
    =\{ f_{D3}(\mu_s^2, \mu_L^2)-f_3 (Q^2, \mu_L^2)+f_{D4}(\mu_v^2,\mu_L^2)-f_4(Q^2,\mu_L^2) \} 
$$
$$
    - \{ f_{D3}(\mu_s^2, \mu_L^{\prime 2} )-f_3 (Q^2, \mu_L^{\prime 2})+f_{D4}(\mu_v^2, \mu_L^{\prime 2} )
-f_4(Q^2, \mu_L^{\prime 2}  ) \} 
 \eqno{(40)} 
$$
and 
$$  \Delta f_t(\Lambda^2>\vert k^2\vert >\Lambda^{\prime 2})
    =f_t (Q^2,\mu_L^2 )-f_t(Q^2,\mu_L^{\prime 2} )
$$
$$
    =-\{ f_5 (Q^2,\mu_L^2)+f_6(Q^2,\mu_L^2) \} 
    +\{f_5 (Q^2,\mu_L^{\prime 2} )+f_6(Q^2,\mu_L^{\prime 2} ) \} 
, \eqno{(41)} 
$$
where $\mu_L^\prime =\Lambda^\prime /M$. 
It is easy to see that $f_i(Q^2,\mu_L^{\prime 2} )$, $f_{Di}(\mu_s^2,\mu_L^{\prime 2})$ and $f_{Di}(\mu_v^2,\mu_L^{\prime 2})$ are also the functions of $M^2/\Lambda^{\prime 2}$ and $-q^2/\Lambda^{\prime 2}$. 
For an example, the original form of $f_4(Q^2,\mu_L^{\prime 2})$ is 
$$
f_4(Q^2,\mu_L^{\prime 2})=-{{g_v^2}\over{8\pi^2}}\int_0^1 dx\int_0^{1-x}dy
[{{2(1-x-y)-(x+y)^2+Q^2(1-x)(1-y)}\over{(x+y)^2+xyQ^2+\mu_L^{\prime 2}(1-x-y)}}
$$
$$
+\ln{\big( {{(x+y)^2+xyQ^2+\mu_L^{\prime 2}(1-x-y)}\over{(x+y)^2+\mu_L^{\prime 2}(1-x-y)}}\big)}]. \eqno{(42)} $$
This equation can be transformed into 
$$
f_4(Q^2,\mu_L^{\prime 2})=-{{g_v^2}\over{8\pi^2}}\int_0^1 dx\int_0^{1-x}dy
[{{2(1-x-y)A-(x+y)^2A+B(1-x)(1-y)}\over{(x+y)^2A+xyB+(1-x-y)}}
$$
$$
+\ln{\big( {{(x+y)^2A+xyB+(1-x-y)}\over{(x+y)^2A+(1-x-y)}}\big)}]
$$
$$
\equiv {\tilde{f}}_4(A,B), \eqno{(43)} 
$$
where $A=M^2/\Lambda^{\prime 2}$ and $B=-q^2/\Lambda^{\prime 2}$ and we use the mark "tilde" to denote that the function with it is the function of $A$ and $B$. 
In a similar way, the function $f_6(Q^2,\mu_L^{\prime 2} )$ is also written in the form
$$ f_6(Q^2,\mu_L^{\prime 2} )
={{g_v^2}\over{8\pi^2}}\int_0^1 dx\int_0^{1-x}dy
{{A(1-x-y)(x+y)}\over{(x+y)^2A+xyB+(1-x-y)}} $$
$$
\equiv {\tilde{f}}_6(A,B), \eqno{(44)} 
$$
Furthermore, $f_i(Q^2,\mu_L^2)$, $f_{Di}(\mu_s^2, \mu_L^2)$ and $f_{Di}(\mu_v^2,\mu_L^2)$ are the function of $A$, $B$ and $E=\Lambda^\prime/\Lambda$. 
Therefore, $\Delta f$ is the function of $A$, $B$ and $E$. 
We denote them as $\Delta {\tilde{f}}(A,B;E)$. 
If $\vert q^2 \vert$ and $M$ are much smaller than $\Lambda^{\prime 2}$, 
we can expand $\Delta {\tilde{f}}$ in the powers of $M^2/\Lambda^{\prime 2}$ and $-q^2/\Lambda^{\prime 2}$ treating $E$ as of order 1, i.e., 
$$ \Delta {\tilde{f}}_s=\sum_{l=0}^\infty\sum_{m=0}^\infty C_{s,l,m}
\big( {{M^2}\over{\Lambda^{\prime 2}}} \big)^l
\big( {{-q^2}\over{\Lambda^{\prime 2}}} \big)^m
$$
$$
\equiv\sum_{n=0}^\infty {{1}\over{n!}}
\big( {{M^2}\over{\Lambda^{\prime 2}}}{{\partial}\over{\partial A}}+
{{-q^2}\over{\Lambda^{\prime 2}}}{{\partial}\over{\partial B}}\big)^n
\Delta {\tilde{f}}_s(A,B;E)\vert_{A=0,B=0}, \eqno{(45)} $$
$$ \Delta {\tilde{f}}_v=\sum_{l=0}^\infty\sum_{m=0}^\infty C_{v,l,m}
\big( {{M^2}\over{\Lambda^{\prime 2}}} \big)^l
\big( {{-q^2}\over{\Lambda^{\prime 2}}} \big)^m
$$
$$
\equiv\sum_{n=0}^\infty {{1}\over{n!}}
\big( {{M^2}\over{\Lambda^{\prime 2}}}{{\partial}\over{\partial A}}+
{{-q^2}\over{\Lambda^{\prime 2}}}{{\partial}\over{\partial B}}\big)^n
\Delta {\tilde{f}}_v(A,B;E)\vert_{A=0,B=0} \eqno{(46)} $$
and
$$ \Delta {\tilde{f}}_t=\sum_{l=0}^\infty\sum_{m=0}^\infty C_{t,l,m}
\big( {{M^2}\over{\Lambda^{\prime 2}}} \big)^l
\big( {{-q^2}\over{\Lambda^{\prime 2}}} \big)^m
$$
$$
\equiv \sum_{n=0}^\infty {{1}\over{n!}}
\big( {{M^2}\over{\Lambda^{\prime 2}}}{{\partial}\over{\partial A}}+
{{-q^2}\over{\Lambda^{\prime 2}}}{{\partial}\over{\partial B}}\big)^n
\Delta {\tilde{f}}_t(A,B;E)\vert_{A=0,B=0}. \eqno{(47)} $$
As is in the case of $f_4(Q^2,\mu_L^{\prime 2})$ (see eq. (43)) and $f_6(Q^2,\mu^{\prime 2})$ (see eq. (44)), $f_i(Q^2,\mu_L^{\prime 2})$ ( $f_i(Q^2,\mu_L^2)$ ) are of order $1/\Lambda^{\prime 2}$ ($1/\Lambda^2$). 
In particular, the leading terms of $f_5(Q^2,\mu^{\prime 2})$ and $f_6(Q^2,\mu^{\prime 2})$ are of order $M^2/\Lambda^{\prime 2}$ and the leading terms do not have $Q$-dependence. 
Therefore, $\Delta f_t$ is of order $M^2/\Lambda^{\prime 2}$, i.e., $C_{t,0,0}=0$ and $C_{t,0,1}=0$. 
On the other hand, $f_{Di}(\mu_s^2,\mu_L^{\prime 2})$ ($f_{Di}(\mu_s^2,\mu_L^2)$) and $f_{Di}(\mu_v^2,\mu_L^{\prime 2})$ ($f_{Di}(\mu_v^2,\mu_L^2)$) have logarithmic behaviors of $\Lambda^\prime$ ($\Lambda$). 
Therefore, $C_{s,0,0}$ and  $C_{v,0,0}$ have the logarithmic behavior of $E=\Lambda^\prime /\Lambda$. 

The contributions of $\Delta {\tilde{f}}$ cannot be dropped, since the cutoff $\Lambda^\prime$ are just conveniently introduced and has no physical meaning. 
However such a contribution can be reincorporated into the theory, by adding the new interaction $\Delta L$ to $L$. 
For an example, up to the order $1/\Lambda^{\prime 2}$, $\Delta L$ is given as follows. 
$$\Delta L=
{\bar{\psi}}g_s\big( C_{s,0,0}+C_{s,1,0}{{M^2}\over{\Lambda^{\prime 2}}}\big) \phi \psi
-{\bar{\psi}}g_v\big( C_{v,0,0}+C_{v,1,0}{{M^2}\over{\Lambda^{\prime 2}}}\big) \gamma_{\mu}V^\mu\psi$$
$$+{{g_vC_{t,1,0}M}\over{2 \Lambda^{\prime 2}}}{\bar{\psi}}F^{\mu\nu}\sigma_{\mu\nu} \psi$$
$$+{{g_sC_{s,0,1}}\over{\Lambda^{\prime 2}}}{\bar{\psi}}\partial_\mu\partial^\mu \phi \psi $$
$$-{{g_vC_{v,0,1}}\over{\Lambda^{\prime 2}}}{\bar{\psi}}\partial_\mu F^{\mu\nu}\gamma_{\nu}\psi \eqno{(48)} $$
The first two terms of eq. (48) have logarithmic behaviors $constant\times \log{\big( \Lambda^\prime /\Lambda\big)}$ in the large $\Lambda^\prime$ limit. 
However they can be absorbed in the redefinitions of the counter term $\delta L_{vertex}$, i.e.,
$$ \delta L^\prime_{vertex} =-{\bar{\psi}}g_s\big( \zeta_s-C_{s,0,0}-C_{s,1,0}{{M^2}\over{\Lambda^{\prime 2}}}\big) \phi \psi
+{\bar{\psi}}g_v\big( \zeta-C_{v,0,0}-C_{v,1,0}{{M^2}\over{\Lambda^{\prime 2}}}\big) \gamma_{\mu}V^\mu\psi$$
$$ \equiv -{\bar{\psi}}g_s\zeta_s^\prime\phi \psi
+{\bar{\psi}}g_v\zeta^\prime \gamma_{\mu}V^\mu\psi \eqno{(49)} $$
is the new counter term which is used with the cutoff $\Lambda^\prime$. 
This change does not affect the physical results, if, instead of $\zeta_s$ and $\zeta_v$, $\zeta_s^\prime$ and $\zeta_v^\prime$ are determined to make $g_s$ and $g_v$ be the physical observed values of the $\sigma$-nucleon and $\omega$-nucleon coupling respectively. 
The last three terms of eq. (48) are of order $(1/\Lambda^{\prime 2})$ and essentially new terms. 
Adding $\Delta L$ to the original $L$, we can get the results the errors of which is of order $(1/\Lambda^\prime)^4$, if we use the $\Lambda^\prime$ instead of $\Lambda$. 
If the higher accuracy is needed, the higher terms which correspond to the higher terms in the expansions (45)$\sim$(47) should be added to $\Delta L$. 

The discussions above is essentially the same as in the case of QED [11]. 
However, it should be remarked that, different from the QED case, the nucleon mass $M$ may not be much smaller than $\Lambda^\prime$. 
Therefore, we formally regard $M$ as the same order of $\Lambda^\prime$. 
The expansions in the powers of $-q^2/\Lambda^{\prime 2}$ are essentially equivalent to the expansions in the powers of $-q^2/M^2=Q^2$, when $M$ is the same order of $\Lambda^\prime$. 
In this respect, $\Delta f_s$, $\Delta f_v$ and $\Delta f_t$ are reexpanded as follows. 
$$\Delta f_s=\sum_{m=0}^\infty \kappa_{s,m}(Q^2)^m=\sum_{m=0}^\infty {{1}\over{m!}}
{{\partial^m ( \Delta f_s)}\over{\partial (Q^2)^m}}\vert_{Q^2=0}(Q^2)^m, \eqno{(50)} $$
$$
\Delta f_v=\sum_{m=0}^\infty \kappa_{v,m}(Q^2)^m=\sum_{m=0}^\infty {{1}\over{m!}}
{{\partial^m (\Delta f_v) }\over{\partial (Q^2)^m}}\vert_{Q^2=0}(Q^2)^m, \eqno{(51)} $$
and 
$$
\Delta f_t=\sum_{m=0}^\infty \kappa_{t,m}(Q^2)^m=\sum_{m=0}^\infty {{1}\over{m!}}
{{\partial^m (\Delta f_t)}\over{\partial (Q^2)^m}}\vert_{Q^2=0}(Q^2)^m. \eqno{(52)} $$
The factor ${{1}\over{2}}$ in the definition of the expansion of the tensor part makes one to compare the zero-th order tensor coupling $\kappa_{t,0}$ directly with the one which is defined usually. 
If we want the results the errors of which is of order $1/\Lambda^{\prime 4}$, we must add 
$$\Delta L^*=
{\bar{\psi}}g_s\kappa_{s,0}\phi \psi
-{\bar{\psi}}g_v\kappa_{v,0}\gamma_{\mu}V^\mu\psi$$
$$+{{g_v\kappa_{t,0}}\over{4M}}{\bar{\psi}}F^{\mu\nu}\sigma_{\mu\nu} \psi $$
$$+{{g_s\kappa_{s,1}}\over{M^2}}
{\bar{\psi}}\partial_\mu\partial^\mu \phi \psi $$
$$-{{g_v\kappa_{v,1}}\over{M^2}}{\bar{\psi}}\partial_\mu F^{\mu\nu}\gamma_{\nu}\psi \eqno{(53)} $$
to the original Lagrangian (1). 
The expression of $\Delta L^*$ resembles $\Delta L$, however, different from $\Delta L$, the each term of $\Delta L^*$ ( eq. (53)) has the contributions of the all higher order of $M^2/\Lambda^{\prime 2}$. 
The new coupling constants $\kappa_{t,0}$, $\kappa_{s,1}$ and $\kappa_{v,1}$ depend on the values of $\Lambda$ and $\Lambda^\prime$. 
In fig. 4, we show the $\Lambda^\prime$-dependence of these couplings in the case of $\Lambda =2$GeV. 
The coupling constants $\kappa_{s,0}$, $\kappa_{v,0}$ and $\kappa_{t,1}$ are 
also shown for comparison. 
The couplings $\kappa_{s,0}$ and $\kappa_{v,0}$ correspond to the changes of the bare scalar and the bare vector couplings in the Lagrangian when the value of the cutoff is changed. 
When $\Lambda^\prime =1$GeV, they are not small. 
However, they does not change the physical results, if the bare couplings are chosen phenomenologically. 
The couplings $\kappa_{s,1}$ and  $\kappa_{v,1}$ are small. 
Furthermore, their contributions are suppressed by the factor 
$-q^2/M^2$. 
The coupling $\kappa_{t,1}$ is very small. 
Its contribution is suppressed by the factor 
$(-q^2)^{3/2}/M^3$. 
On the other hand, $\kappa_{t,0}\sim 0.4$ in the case of $\Lambda^\prime=1$GeV. 
The weak tensor coupling may be needed if the difference between 
$\Lambda$ and $\Lambda^\prime$ is order of GeV. 


\centerline{$\underline{~~~~~~~}$}

\centerline{Figs. 4 (a),(b),(c)}

\centerline{$\underline{~~~~~~~}$}


In figure 5, we show the contributions of these couplings in the case of $\Lambda=2$GeV and $\Lambda^\prime =1$GeV. 
Although we are interested in the small $Q^\prime (=\sqrt{-q^2/\Lambda^{\prime 2}}<1)$ region, we show the result in the region $Q^\prime =0\sim 2$ to see the $Q^\prime$-dependence of the efficiency of the approximation. 
In the cases of $F_s$ and $F_v$, the contributions which are discarded by changing the cutoff $\Lambda$ to $\Lambda^\prime$ are well reincorporated 
in the small $Q^\prime (<1)$ region by introducing the new term (53). 
In these figures, we see that $\kappa_{s,1}$ and $\kappa_{v,1}$ are important to see the momentum dependence of the vertex corrections even in the region $Q^\prime <1$, although they are small. 
In the case of $F_t$, the result with the correction (53) seems to deviate somewhat from the true result (solid line). 
However, it should be remarked that the contribution of $F_t$ itself is suppressed by the factor $\sqrt{-q^2}/M$. 
Therefore, the difference could be negligible 
in the small $q$ region. 
If the higher term with the coefficient $\kappa_{t,1}$ is taken into account, 
the result hardly deviates from the true result (solid line) in the small $Q^\prime (<1)$ region. 

As we have seen above, the new interaction in (53) is very important when the 
difference between $\Lambda$ and $\Lambda^\prime$ is order of GeV. 
It is also shown that the $Q$ expansion have a good convergence even if $\Lambda^\prime$ is not so larger than the nucleon mass $M$. 
This situation is the same as the one in the relativistic Hartree calculation of the energy density of the nuclear matter [8].


\centerline{$\underline{~~~~~~~}$}

\centerline{Figs. 5 (a), (b), (c)}

\centerline{$\underline{~~~~~~~}$}


\centerline{\bf 4. Summary and Discussions}

\bigskip

The results obtained in this paper are summarized as follows. 

(1) At zero density, we have studied the vertex corrections in the cutoff field theory based on the $\sigma$-$\omega$ model. 
It is shown that the results depend strongly on the value of the cutoff $\Lambda$. 
The scalar and vector vertex functions decrease as $\Lambda$ increases. 
On the other hand, the tensor vertex functions increases as $\Lambda$ increases. 

(2) The results in the cutoff field theory was compared with the results obtained by the ordinary renormalization procedures ( the case with $\Lambda\rightarrow\infty$) . 
If $\Lambda$ is large ( $\sim$20GeV ), the former results are the almost same as the latter one. 
On the other hand, the results are somewhat different from those in the ordinary renormalization procedures if $\Lambda$ is small ($<$5GeV). 

(3) The cutoff field theory is reformulated in the framework of the renormalization group methods by introducing the convenient cutoff $\Lambda^\prime$ into the theory. 
Instead of the naive expansion of $1/\Lambda^{\prime 2}$, regarding that $M$ is the same order of $\Lambda^\prime$, $-q^2/M^2$ expansion is adopted to construct the low-energy effective Lagrangian in QHD. 
It is shown that the expansion has a good convergence even if $\Lambda^\prime$ is not so much larger than the nucleon mass $M$. 

(4) The effects of the contributions which are needlessly discarded by the using the smaller cutoff $\Lambda^\prime$ instead of the true cutoff $\Lambda$ are well reincorporated into the theory with the new interactions up to the order $1/\Lambda^{\prime 2}$. 

(5) It seems that the tensor coupling and the derivative couplings are needed to be added to the Lagrangian when the difference between the convenient cuf-off $\Lambda^\prime$ and the true cutoff $\Lambda$ is order of GeV. 

In this paper, we have estimated the couplings of the low-energy effective Lagrangian with the assumption of $\Lambda$ and have studied the effects of the new interaction terms. 
However, if we use the vertex functions in more complicated models such as in the nuclear Schwinger-Dyson formalism [19], it may be more convenient to determine the couplings in the new interaction 
$$\Delta L^{* \prime }=
{{g_v\kappa_{t,0}}\over{4M}}{\bar{\psi}}F^{\mu\nu}\sigma_{\mu\nu} \psi $$
$$+{{g_s\kappa_{s,1}}\over{M^2}}
{\bar{\psi}}\partial_\mu\partial^\mu \phi \psi $$
$$-{{g_v\kappa_{v,1}}\over{M^2}}{\bar{\psi}}\partial_\mu F^{\mu\nu}\gamma_{\nu}\psi \eqno{(54)} $$
phenomenologically as well as the coefficients $\zeta_s^\prime$ and $\zeta_v^\prime$ in the new counterterm $\delta L^\prime_{vertex}$ [11]. 
The main merits of this phenomenological method are following. 

(1) The effects of the corrections beyond the approximations we use are also taken into account by the phenomenological fits as well as the effects which are needlessly discarded by using the small cutoff $\Lambda^\prime$ instead of the true cutoff $\Lambda$. 

(2) In this phenomenological fits, $\Lambda^\prime$-dependence of the results is removed by adding new interactions to the Lagrangian, order by order. 
Of course, the dependence on the true cutoff $\Lambda$ exists, since it has a physical meaning. 
However, it appears only in the coefficients of $\delta L^\prime$ and $\Delta L^{* \prime}$ which are phenomenologically determined. 
Therefore, the $\Lambda$-dependence does not appear explicitly in the theory after those phenomenological fits are done. 
It should be also remarked that, by the following reason, the results hardly depend on the details of the regulator. 
The fundamental forms of the expansions (54) does not change if we use the other type of regulator instead of the Pauli-Villars regulator. 
It is determined just by the power counting of $q^\mu/M$, e.g., if we need the results errors of which are of order $1/\Lambda^{\prime 4}$, we just need to add the new interaction terms which correspond to the $-q^2/M^2$ terms in the $Q$-expansions of the scalar and vector vertex corrections and add the new interaction term which corresponds to $q^\mu/M$ term in the $Q$-expansion of the tensor vertex correction, respectively. 
They are written in the same forms as in eq. (54). 
In general, if we want the results the errors of which is of order $(1/\Lambda^{\prime 2})^{n+1}$, we need to add the new terms which correspond to $(-q^2/M^2)^m$ terms ($m\leq n$, where $n$ and $m$ are integers ) in the $Q$-expansions of the scalar and vector vertex corrections, and need to add the new terms which correspond to $(q^\mu/M)(-q^2/M)^{m-1}$ terms ($m\leq n$) in the $Q$-expansion of the tensor vertex correction. 
The coupling constants of those terms depend on the details of the regulator which we use. 
However, they are determined phenomenologically. 
Therefore, the physical results does not depend on the details of the regulator we use as well as on the value of $\Lambda^\prime$, except for the errors of order $(-q^2/\Lambda^{\prime 2})^{n+1}$. 
This is a nice merit, since, in the naive cutoff theory such as the one in the section 2, the results may depend on the details of the regulator we use even if the value of $\Lambda$ is determined phenomenologically. 

(3) Phenomenological approach may be very useful in the actual numerical calculations, since we do not need to calculate the momentum dependence of the physical quantities in the high momentum region ($\vert q^2 \vert > \Lambda^{\prime 2}$) in which we are not interested. 

It should be also remarked that the difference between $\Delta L$ and $\Delta L^*$ makes no difference in the physical results, since all the coefficients of them are determined phenomenologically. 

Finally we add the following comment. 
Instead of starting at the original Lagrangian $L$ with the original cutoff $\Lambda$, we can also start at the Lagrangian $L^\prime=L+\Delta L$ (or $L +\Delta L^*$) with the cutoff $\Lambda^\prime$, introducing the cutoff $\Lambda^{\prime\prime}$ which is smaller than $\Lambda^\prime$, and get the effective Lagrangian $L^{\prime\prime}$ at lower energy. 
The resulting effective Lagrangian $L^{\prime\prime}$ is almost the same as the one obtained directly from the original Lagrangian $L$ in a similar way, except for the errors of order $1/\Lambda^{\prime 4}$. 
This mean that, in the cutoff field theory, the Lagrangian with the tensor and derivative couplings such as in $\Delta L$ or $\Delta L^*$ has no difficulty as the "renormalizable" Lagrangian such as eq. (1), if the tensor and derivative couplings are suppressed by the factor of order $1/\Lambda^2 \sim 1/\Lambda^{\prime 2}$. 
On the other hand, if those couplings are order of 1 or larger, they generate the Feynman diagrams of order $\Lambda^{\infty}$ and it is impossible to construct the effective Lagrangian at lower energy scale within finite terms, even if the errors of some finite order of $1/\Lambda^{\prime 2}$ are allowed. 
This situation corresponds to the "renormalizability" in the limit of $\Lambda^\prime \rightarrow \infty$. 

In this paper, we restrict our discussions to the zero baryon density. 
It is very interesting to generalize this result to the case at finite baryon density and to calculate the energy density by including the effects of the vertex corrections. 
They are now under the studies. 

\noindent
Acknowledgement: The authors gratefully thank K. Harada, who have recommended them to use the renormalization group methods in quantum hadrodynamics, for many useful suggestions and discussions. 
They also thank T. Kohmura, T. Suzuki, M. Yahiro and H. Yoneyama for useful discussions, 
 and acknowledge the computing time granted by Research Center for Nuclear Physics (RCNP). 

\vfill\eject

\centerline{\bf References}


\noindent
[1] J.D. Walecka, Ann. of Phys. {\bf 83} (1974) 491. 

\noindent
[2] S.A. Chin, Phys. Lett. {\bf 62B} (1976) 263. 

\noindent
[3] S.A. Chin, Ann. of Phys. {\bf 108} (1977) 301

\noindent
[4] T. Kohmura, Y. Miyama, T. Nagai, S. Ohnaka, J. Da Provid{\^{e}}ncia, and T. Kodama, Phys. Lett. {\bf B226}(1989)207. 

\noindent
[5] R.J. Furnstahl and C.J. Horowitz, Nucl. Phys. {\bf A485} (1988) 632.

\noindent
[6] H. Kurasawa and T. Suzuki, Nucl. Phys. {\bf A445} (1985) 685: 

\noindent
H. Kurasawa and T. Suzuki, Nucl. Phys. {\bf A490} (1988) 571. 

\noindent
[7] T.D. Cohen, Phys. Lett. {\bf B211}(1988)384. 

\noindent
[8] H. Kouno, T. Mitsumori, Y. Iwasaki, K. Sakamoto, N. Noda, K. Koide, A. Hasegawa and M. Nakano, Prog. Theor. Phys., {\bf 97} (1997)91. 

\noindent
[9] K.G. Wilson and J. Kogut, Phys. Rep. {\bf 12} (1974)75. 

\noindent
[10] K. G. Wilson, Rev. Mod. Phys. {\bf 55} (1983)583. 

\noindent
[11] G P. Lepage, "What is renormalization?" in "From actions to answers", Proceedings of the 1989 theoretical advanced study institute in elementary particle physics, edited by T. DeGrand and D. Toussaint, p483, (World Scientific, Singapore 1990). 

\noindent
[12] H. Kouno, K. Sakamoto, Y. Iwasaki, N. Noda, T. Mitsumori, K. Koide, A. Hasegawa and M. Nakano, Preprint SAGA-HE-120-97 (nucl-th/9703055). 

\noindent
[13] M. Nakano, N. Noda, T. Mitsumori, K. Koide, H. Kouno and A. Hasegawa, 
Phys. Rev. {\bf C55} (1997) 890. 

\noindent
[14] M. P. Allendes and B. D. Serot, Phys. Rev. {\bf C45} (1992) 2975.

\noindent
[15] B. D. Serot and H.-B. Tang, Phys. Rev. {\bf C51} (1995) 969.

\noindent
[16] W. Pauli and F. Villars, Rev. Mod. Phys. {\bf 21} (1949) 434. 

\noindent
[17] N. N. Bogoliubov and D. V. Shirkov, "Introduction to the theory of quantized fields", Interscience publisher, New York, 1959 : 
N. N. Bogoliubov and D. V. Shirkov, "Quantum fields", The Benjamin/Cummings Publishing Company, Canada, 1983

\noindent
[18] C. Itzykson and J-B. Zuber, "Quantum field theory", McGraw-Hill, New York, 1980. 

\noindent

\noindent
[19] M. Nakano, A. Hasegawa, H. Kouno, and K. Koide, Phys. Rev. {\bf C49} (1994) 3061. 


\vfill\eject

\centerline{\bf Figure Captions}

\bigskip

\bigskip

\noindent
Fig. 1 Diagrammatic expansion of the $NN\sigma$ or $NN\omega$ vertexes. 
The solid line represents the nucleon and the dotted line represents the meson. 

\bigskip

\bigskip

\noindent
Fig. 2 The $Q$-dependence of the vertex corrections in the cases of $\Lambda =1$GeV (dashed line) and 2GeV (dotted line). 
The result obtained by the ordinary renormalization procedures ( in the case of $\Lambda =\infty$ ) is also shown (solid line) for comparison. 
(a) Scalar vertex. 
(b) Vector vertex. 
(c) Tensor vertex. 

\bigskip

\bigskip

\noindent
Fig. 3 The $\Lambda$-dependence of the vertex corrections in the case of $QM=0.5$GeV (dotted line). 
The result obtained by the ordinary renormalization procedures ( in the case of $\Lambda =\infty$ )which is not the function of $\Lambda$ is also shown by the solid line for comparison. 
(a) Scalar vertex. 
(b) Vector vertex. 
(c) Tensor vertex. 

\bigskip

\bigskip

\noindent
Fig. 4 The $\Lambda^\prime$-dependence of the coefficients of the expansion of eqs. (50)$\sim$(52). 
(a) Scalar vertex. 
The solid line and the dotted line are $\kappa_{s,0}$ and $\kappa_{s,1}$, respectively. 
(b) Vector vertex. 
The solid line and the dotted line are $\kappa_{v,0}$ and $\kappa_{v,1}$, respectively. 
(c) Tensor vertex. 
The solid line and the dotted line are $\kappa_{t,0}$ and$\kappa_{t,1}$, respectively. 

\bigskip

\bigskip

\noindent
Fig. 5 The $Q^\prime$-dependence of the vertex corrections in the case of $\Lambda =$2GeV. 
(a) Scalar vertex. 
(b) Vector vertex. 
(c) Tensor vertex. 
The solid line is the true result at $\Lambda^\prime =\Lambda =2$GeV. 
The dotted line is the result at $\Lambda^\prime =1$GeV without the tensor and derivative coupling terms. 
The dashed line is the result at $\Lambda^\prime =1$GeV with the new interaction term which is proportional to $\kappa_{s,1}$ ($\kappa_{v,1}$, $\kappa_{t,0}$) in Fig. 5(a) (Fig. 5(b), Fig. 5(c)). 
In Fig. 5(c), the result with the higher order term with the coefficient $\kappa_{t,1}$ is also shown for comparison (dashed-dotted line).

\end{document}